\documentstyle[12pt,aaspp4]{article}

\def\ltsima{$\;\buildrel < \over \sim \;$}
\def\simlt{\lower.5ex \hbox{\ltsima}}
\def\gtsima{$\;\buildrel > \over \sim \;$}
\def\simgt{\lower.5ex \hbox{\gtsima}}
\begin{document}
\title{Discovery of Interstellar Hydrogen Fluoride$^*$}

\author{David A. Neufeld$^1$, Jonas Zmuidzinas$^2$, Peter Schilke$^3$ 
and Thomas G. Phillips$^2$}

\vskip 0.5 true in
{\parskip 6pt
\noindent{$^1$Department of Physics \& Astronomy, 
The Johns Hopkins University, 3400 North Charles Street, 
Baltimore, MD 21218}

\noindent{$^2$Division of Mathematics, Physics, \& Astronomy,
320-47, California Institute of Technology, Pasadena, CA 91125}

\noindent{$^3$Max-Planck Institute for Radio Astronomy, 
Auf dem H\"ugel 69, D-53121, Bonn, Germany}
\vskip 0.5 true in
\noindent{$^*$ Based on observations with ISO, an ESA project with instruments
     funded by ESA Member States (especially the PI countries: France,
     Germany, the Netherlands and the United Kingdom) with the
     participation of ISAS and NASA.}
\vskip 0.5 true in}


\begin{abstract}

We report the first detection of interstellar hydrogen fluoride.
Using the Long Wavelength Spectrometer (LWS) of the Infrared
Space Observatory (ISO), we have detected the 121.6973 $\mu$m
$J=2-1$ line of HF in absorption toward the far-infrared
continuum source Sagittarius B2.  The detection is statistically
significant at the $13 \,\sigma$ level.   On the basis of our model for
the excitation of HF in Sgr B2, the observed line equivalent 
width of $1.0$ nm implies a hydrogen fluoride abundance 
of $\sim 3 \times 10^{-10}$ relative to H$_2$.  If the elemental
abundance of fluorine in Sgr B2 is the same as that in the solar 
system, then HF accounts for $\sim 2 \%$ of the total number of
fluorine nuclei.  
We expect hydrogen fluoride to be the dominant reservoir of
gas-phase fluorine in Sgr B2, because it is formed rapidly in 
exothermic reactions of atomic fluorine with either water or 
molecular hydrogen; thus the measured HF abundance suggests a 
substantial depletion of fluorine onto dust grains.  
Similar conclusions regarding depletion
have previously been reached for the case of chlorine in dense interstellar
clouds.  We also find evidence at a lower level of statistical significance
($\sim 5\,\sigma$) for an emission feature at the expected position 
of the $4_{32}-4_{23}$ 121.7219~$\mu$m line of water.  The emission 
line equivalent width of $0.5$~nm for the water feature is
consistent with the water abundance of
$5 \times 10^{-6}$ relative to H$_2$ that has been inferred 
previously from observations of the hot core of Sgr B2.

\end{abstract}

\section{Introduction}

Molecules are a ubiquitous component of the dense interstellar 
medium.  To date, about one hundred distinct species have been detected in
the interstellar gas (Ohishi 1997); they range from simple diatomic molecules
to complex species containing as many as thirteen atoms.  The wide variety 
of interstellar molecules demonstrates the thermodynamic tendency 
of most elements to form molecules under the conditions present in the
dense interstellar medium.  Although most of the interstellar molecules
that have been detected previously contain elements\footnote{viz.\ 
hydrogen, oxygen, carbon, nitrogen, sulphur and silicon}
of cosmic abundance
greater than $10^{-5}$, a few 
molecules containing elements
of lower abundance have also been detected.  For example, chlorine, 
with a solar system
abundance of only 2$\times 10^{-7}$ relative to hydrogen
(Anders \& Grevesse 1989), was detected 
more than a decade ago in the form of 
hydrogen chloride (Blake, Keene \& Phillips 1985).

Recent estimates (Neufeld \& Green 1994; Schilke, Phillips \& Wang
1995; Zmuidzinas et al.\ 1995a) of the HCl abundances 
implied by observations of the HCl $J=1-0$ line toward Sgr B2
(Zmuidzinas et al.\ 1995a) and the Orion Molecular Cloud 1 
(Blake et al.\ 1985;
Schilke et al.\ 1995) have yielded results for 
$n({\rm HCl})/n({\rm H}_2)$ in the range 0.3 -- 2 $\times 10^{-9}$.   
If the elemental abundance of chlorine in these 
sources is the same as that in 
the solar system, then HCl accounts for only $0.1 - 0.7\%$ of the 
total number of chlorine nuclei.  Theoretical models for the
chemistry of chlorine-bearing molecules (Schilke et al.\ 1995)
predict that HCl will account for $\sim 30\%$ of gas-phase chlorine.
Thus the observed abundance of HCl can only be understood if the chlorine 
depletion in the dense ISM is large.  The required depletion factors
of $\simgt 10^2$ greatly exceed the values inferred for the {\it diffuse}
ISM from UV absorption line studies (e.g.\ Harris, Gry, \& Bromage 1984 ).
       
Prior to the observations reported in this
Letter, HCl was the only known
{\it interstellar} molecule containing a halogen element\footnote
{Note, however, that NaCl, KCl, AlCl, and AlF have been detected 
in the circumstellar envelope of IRC+10216 (Cernicharo \& Gu\'elin 1987;
Ziurys, Apponi \& Phillips 1994), and that HF lines
have been widely observed in the spectra of cool stars (e.g. Jorissen,
Smith \& Lambert 1992).}.  
Motivated by the earlier observations of HCl, by the fact
that the solar system abundance of fluorine lies only a factor
of 6 below that of chlorine (Anders \& Grevesse 1989), and by
the large H--F bond strength which suggests that hydrogen fluoride is likely
to be a major reservoir of gas-phase fluorine,
we have  undertaken a search 
for interstellar hydrogen fluoride toward the strong far-infrared
continuum source Sgr B2.  Such a search would provide a unique
probe of the chemistry of interstellar fluorine and of the
fluorine depletion in the dense interstellar medium.
Because of its large rotational constant --
the largest of any diatomic molecule other than molecular hydrogen 
or HeH$^+$ --
HF possesses a rotational spectrum that lies entirely shortward of
the atmospheric windows within which most molecules show a rotational
spectrum.  Observations of HF rotational transitions are therefore
possible only from airplane or space-based observatories.  We have
made use of the Long Wavelength Spectrometer (LWS; 
Clegg et al.\ 1996) on board the 
Infrared Space Observatory (ISO; Kessler et al.\
1996) to search for hydrogen fluoride.
The $J=1-0$ transition of HF lies longward of the wavelength range
to which the photoconductive detectors used in LWS are sensitive, so 
we have carried out observations of the $J=2-1$ line at 121.6973 $\mu$m.
These observations are described in \S 2 below.  Our results
are presented in \S 3, and discussed in \S 4.  The detection of
hydrogen fluoride reported here marks the first discovery of an
{\it interstellar} molecule containing the element fluorine and the
first time that a new astrophysical molecule has been identified
by means of observations in the far-infrared (30 -- 300 $\mu$m) 
spectral region.

\section{Observations and data reduction}

Using the LWS of ISO in Fabry-Perot mode at a 
resolving power $\lambda/\Delta \lambda$ of 9600, 
we observed the
$J=2-1$ line of HF toward the source Sgr B2 on
1997 March 28th.  The rest
frequency of the line is $\rm 2\,463\,428.11 \pm 0.21\,\rm MHz$ (Nolt et
al.\ 1987),
corresponding to a vacuum wavelength of 121.6973~$\mu$m.
Unlike HCl, HF shows no electric quadrupole hyperfine structure,
and the magnetic hyperfine splitting is far smaller than the
intrinsic width of any astrophysical absorption
line. The ISO beam, of approximate diameter
$70^{\prime\prime}$ FWHM, was centered midway between
Sgr B2(M) and Sgr B2(N), at coordinates $\rm \alpha=17h\,47m\,20.0s$,
$\delta=-28^\circ \,22^\prime\,41.3 ^{\prime\prime}$ (J2000).  
The observation, which covered the wavelength range
121.63 -- 121.80 $\mu$m, was carried out in fast scanning mode
with 8 spectral 
samples per resolution element.  The total integration time
was 2549 seconds.  

The initial data reduction was carried out with version 6.1 of
the ISO pipeline software.  The ISAP software package was then used to
remove bad data points, to co-add the individual spectral scans,
and to derive estimates of the likely statistical error for
each flux measurement from the variance of the individual scans.
The resultant spectrum showed a continuum flux level that varied
by $\sim 13$\% over the narrow wavelength range covered.  This behavior,
which has been widely observed in LWS Fabry-Perot spectra, 
is caused by the grating that is used as an order sorter for the 
Fabry-Perot etalon.  Due to errors in the grating setting, the grating
wavelength is sometimes offset slightly from the central
wavelength of the Fabry-Perot scan, thereby superimposing an 
asymmetric envelope on the measured spectrum
and compromising the absolute flux calibration.  That envelope 
was fit with a second order polynomial, and the absolute flux
calibration was obtained from a spectrum taken in the grating
mode of LWS immediately after the Fabry-Perot scan.

\section{Results}

Figure 1 shows the LWS spectrum observed toward Sgr B2, normalized
with respect to the measured continuum flux of $4.8 \times 10^4$ Jy.  We used
a $\chi^2$ analysis to fit the observed spectrum and
assess the statistical significance of lines that we detected.
Our analysis makes the implicit assumption that the
errors in the fluxes measured
at each spectral point are Gaussian and independent.  We
approximated the spectral response function of the Fabry-Perot instrument
as a Lorentzian with a width corresponding to the nominal resolving power
$\lambda/\Delta \lambda$ of
9600 (Trams, Clegg \& Swinyard 1996); and we 
assumed that the astrophysical
absorption feature was narrow compared to the instrumental
profile, as suggested by the intrinsic line widths measured
for HCl (Zmuidzinas et al.\ 1995a).

Fitting a single absorption feature to the spectrum, and with a 
quadratic fit to continuum as described in \S 2 above, we obtained
an absorption line equivalent width of $1.0$~nm and a central LSR
velocity of 67 km s$^{-1}$ for the 121.6973 $\mu$m HF line.
The measured LSR velocity is in excellent agreement with the 
centroid velocity observed by Zmuidzinas et al.\ (1995a) 
for the HCl $J=1-0$ line in
Sgr B2. Our detection of an absorption feature is statistically 
significant at the $13\,\sigma$ level.

In addition to the absorption feature that we attribute to HF,
we find evidence at a lower level of statistical significance
($\sim 5\,\sigma$) for an emission 
feature near the expected position (for $v_{\rm LSR} =$ 67 km s$^{-1}$)
of the $4_{32}-4_{23}$ line of water, which has a rest wavelength
of 121.7219~$\mu$m (Toth 1991).  The emission
line equivalent width is $0.5$~nm.  With the water emission line
included in the fit, the reduced $\chi^2$ is 0.98 (with 108 degrees of
freedom).

\section{Discussion}

Our identification of the absorption feature in Figure 1 with
HF $J=2-1$ is based upon (1) the excellent agreement of the
observed line position with the expected position of the HF line;
and (2) the absence of any other astrophysically plausible candidate
in available line catalogues.
Notwithstanding these considerations, the detection of additional 
HF lines would be desirable as a confirmation of this identification.
Unfortunately, according to the excitation model discussed below, 
we do not expect absorption lines higher than $J=2-1$ to be
strong enough to be detected by the ISO spectrometers.
Furthermore, the $J=1-0$ transition near 243.2~$\mu$m
lies longward of the spectral region that is observable with the 
photoconductive detectors used in LWS.  
Fortunately, the Earth's atmosphere permits
observations of HF $J=1-0$ from airplane altitude: that
line, with a large predicted equivalent width $\sim 18$~nm,
will therefore be a prime target for observations with
instrumentation for longer wavelengths
that has been proposed for the Stratospheric
Observatory of Infrared Astronomy (SOFIA).
Observations of HF $J=1-0$ will also be possible with
instrumentation proposed for the Far Infrared and Submillimetre
Telescope (FIRST).

In deriving a hydrogen fluoride abundance from the observed 
equivalent width of the $J=2-1$ line, we have made use of an 
excitation model to determine the fractional level populations 
as a function of position within the source.  Our model is essentially
identical to that used by Zmuidzinas et al.\ (1995a) to model the
excitation of HCl in Sgr B2.  The temperature and H$_2$ density in the
source are assumed to vary with the radial coordinate $r=r_{\rm pc}$~pc 
according to 
$$T = 40\,r_{\rm pc}^{-1/2}\,\rm K,$$
$$n({\rm H}_2)/{\rm cm}^{-3} 
= 8.6 \times 10^4\,r_{\rm pc}^{-2} + 2.2 \times 10^{3}.$$
For an assumed dust opacity that varies with wavelength 
as $\lambda^{-1.5}$, these temperature and density profiles
yield a predicted dust continuum spectrum that is in 
excellent agreement with the observations. 
Because of the large dipole moment (1.826 Debye; Muenter 1972) 
and large rotational
constant ($\rm 20.54\,cm^{-1}$) of the HF molecule, the critical
density at which collisional processes become important
in the excitation of HF $J=1$ is very much larger 
($\rm \sim 10^9\,cm^{-3}$) than the density of the gas
($n[{\rm H}_2] \rm \sim 10^5\,cm^{-3}$) 
that we sampled in our observations of Sgr B2.
Thus the excitation of HF is dominated by the effects of
radiative pumping by dust continuum radiation.  
The radiative transfer was treated using the method of 
accelerated lambda iteration (Rybicki \& Hummer 1991),
and the line equivalent width obtained as a function of
the HF abundance.  In the absence of available collisional data for HF,
we adopted as the rate coefficients for collisional de-excitation
those computed previously for the de-excitation of HCl by He
(Neufeld \& Green 1994);  fortunately, because
radiative pumping dominates the excitation of HF
in Sgr B2, our results are almost entirely independent of 
the assumed collisional rate coefficients.  
On the basis of this model, we found that the observed 
equivalent width of $1.0$~nm corresponds to an abundance
$n({\rm HF})/n({\rm H}_2) \sim 3 \times 10^{-10}$.
The predicted equivalent width shows a nearly linear
dependence on the assumed HF abundance (equivalent width
$\propto$ abundance$^{0.85}$) for abundances in the
range $1-6 \times 10^{-10}$.  
Figure 2 shows the fractional level populations in $J=0$, 1, 2
and 3 as a function of radial coordinate for a hydrogen fluoride
abundance of $3 \times 10^{-10}$.  Observations of the $J=2-1$
absorption line sample the region exterior to an effective photosphere
at $r_{\rm pc} \sim 0.6$.

We also modeled the excitation of the H$_2$O $4_{32}-4_{23}$
line that is evident in the spectrum at a lower level
of statistical significance.  For the case of water, 
an excitation model had previously been developed
to account for the observed strengths
(Zmuidzinas 1995b; Gensheimer, Mauersberger \& Wilson 1996)
of the $4_{14}-3_{21}$ and $3_{13}-2_{02}$ 
emission lines of H$_2^{18}$O
and the $1_{10}-1_{01}$ absorption line
of H$_2^{18}$O toward Sgr B2.  The temperature and H$_2$ density
profiles assumed in that model were identical to those
adopted for the HF and HCl excitation models.
To fit the measured fluxes of all three lines, a variable water 
abundance must be assumed, with a smaller water abundance 
$n({\rm H_2^{16}O})/n({\rm H}_2) = 3.3 \times 10^{-7}$ in the cooler
outer parts of the source
($T < 90$~K; $r_{\rm pc} > 0.2$), and a larger water abundance 
of $5 \times 10^{-6}$ in the warmer
inner regions 
(i.e.\ the ``hot core'' with $T \ge 90$~K; $r_{\rm pc} < 0.2$).  
A variation in the water abundance of this nature 
could be explained by the vaporization of water ice from 
grain mantles in the warmer parts of the source.  Water
abundances $\simgt 10^{-5}$ have previously inferred
from observations of H$_2^{18}$O and H$_2^{16}$O
in other warm dense molecular regions (Jacq et al.\ 1988;
Cernicharo et al.\ 1994; Gensheimer et al.\ 1996;
van Dishoeck \& Helmich 1996).
Without the adjustment of any model parameter, our excitation
model for water predicts that
the $4_{32}-4_{23}$ line of H$_2^{16}$O -- 
which is excited primarily within the hot core where the
assumed water abundance is $5 \times 10^{-6}$ -- 
will be an emission line with
an equivalent width of 0.4~nm, in good agreement with
the observations.

To interpret our detection of HF, we have constructed a 
simple model for the steady-state chemistry of interstellar fluorine.
Fluorine is qualitatively different in at least one 
important respect from all the other elements\footnote{viz. oxygen, 
carbon, nitrogen, sulphur, silicon and chlorine} of which 
hydrides have been detected in the interstellar medium: atomic fluorine 
reacts exothermically with H$_2$ via the reaction  
$$\rm H_2 + F  \rightarrow HF + H \phantom{, 00000O}({\rm R1}).$$
Fluorine also undergoes an exothermic reaction with water:
$$\rm H_2O + F  \rightarrow HF + OH\phantom{00000}({\rm R2}).$$
Both reactions have been well-studied in the laboratory, although only at
temperatures greater than 200~K.
For reaction (R1), measurements by three independent groups
have yielded results for the rate coefficient 
at 298 K in the narrow range $\rm 2.3 - 3.0 \times 10^{-11}
cm^3 \,s^{-1}$ (Wurzberg \& Houston 1980; Heidner et al.\ 1980;
Stevens, Brune, \& Anderson 1989).  The temperature dependence of
the measured rate coefficient
implied the presence of a relatively small 
activation energy barrier that
was variously estimated as $E_a/k= 595 \pm 50$, $433 \pm 50$ and 
$470 \pm 30$~K in these three studies.  
We adopted Atkinson et al.'s (1992) review value of 
$k_{\rm H2} = 1.4 \times 10^{-10} \exp\,(-500\,{\rm K}/T)\,\rm cm^3 s^{-1}$
for the rate coefficient for (R1).  For reaction (R2), Stevens et al.\ (1989)
obtained a rate coefficient $k_{\rm H2O} =\rm  (1.6 \pm 0.3) \times 10^{-11}
cm^3 s^{-1}$ with no detectable variation over the temperature range
240--373~K; the implied limit on any activation energy barrier was
($-28 \pm 42$) K.   For a water abundance of
$3.3 \times 10^{-7}$, these rate coefficients
imply that reaction (R1) dominates the formation of HF at temperatures
$\simgt 30$~K.  At the H$_2$ densities present in Sgr B2, both reactions
are expected to dominate ion-neutral reaction networks as a source of HF. 

Once formed, HF is destroyed only very slowly.
HF is more strongly bound than any of the species
CF, CH, OF, OH, NF, NH, SF, SH, SF$^+$, SiF, SiH, or SiF$^+$: thus
reactions of HF with C, O, N, S, S$^+$, Si, 
and Si$^+$ are all endothermic.  HF has a smaller proton affinity
than H$_2$O, CO or N$_2$: thus reactions with H$_3$O$^+$, HCO$^+$,
and $\rm N_2H^+$ are likewise endothermic.  Destruction of HF takes 
place slowly by means of cosmic 
ray induced photodissociation and as a result of reactions with 
species of low abundance such as CH, He$^+$, H$_3^+$, and C$^+$.  

Considering these formation and destruction processes
for hydrogen fluoride, we find that over a wide range of densities 
and temperatures,  including those probed by our observations of 
HF $J=2-1$ in Sgr B2, HF accounts for more than $99\%$ of all fluorine 
nuclei in the gas-phase.  In comparison with the ion-neutral reaction 
routes responsible for the formation of other interstellar hydrides, the 
neutral-neutral reactions (R1) and (R2) are extremely effective 
in producing HF.  Given (1) our theoretical expectation that HF is the 
dominant reservoir of gas phase fluorine along the line-of-sight
to Sgr B2, (2) the assumption that the elemental abundance of
F is the same as that measured in the solar system
($3 \times 10^{-8}$ relative to H; Anders \& Grevesse 1989), and
(3) the observed HF abundance of $\sim 3 \times 10^{-10}$
relative to H$_2$, we conclude that a depletion factor $\sim 50$ 
results from the depletion of fluorine onto dust grains in Sgr B2.  
This is somewhat smaller than the depletion
factors inferred by Zmuidzinas et al.\ (1995a) and Schilke et al.\ (1995) 
for chlorine in Sgr B2 and in OMC-1.
 
We gratefully acknowledge the outstanding support provided by
the Infrared Processing and Analysis Center and particularly
by Steven Lord.  The work of D.A.N. was supported by NASA
grant NAGW-3147 from the Long Term Space Astrophysics Research Program
and by NASA grant NAGW-3183.  J.Z.\ acknowledges support from NASA
contract NAG2-744 and T.G.P.\ from NSF contract AST 96-15025.

\vfill\eject
\centerline{\bf References}
\vskip 0.1 true in 
{\hoffset 20pt
\parindent = -20pt

Anders, E. \& Grevesse, N. 1989, Geochim.\ Cosmochim.\ Acta, 53, 197

Atkinson, R., Baulch, D.L., Cox, R.A., Hampson, R.F. Jr.,
  Kerr, J.A., \& Troe, J. 1992, J.\ Phys. \& Chem.\ Ref.\ Data, 21, 1125

Blake, G.A., Keene, J., \& Phillips, T.G. 1985, ApJ, 295, 501

Cernicharo, J., \& Gu\'elin, M. 1987, A\&A, 183, L10

Cernicharo, J., Gonz\'ales-Alfonso, E., Alcolea, J., Bachiller, R., 
\& John, D. 1994, ApJ, 432, L59.

Clegg, P.E., et al.\ 1996,  A\&A, 315, L38

Gensheimer, P.D., Mauersberger, R., \& Wilson, T.L. 1996, A\&A, 314, 281

%
Harris, A. W., Gry, C., \& Bromage, G. E. 1984, ApJ, 284, 157

Heidner, R.F. III, Bott, J.F., Gardner, C.E., 
  \& Melzer, J.E. 1980, J.\ Chem.\ Phys., 72, 4815


Jacq, T., Henkel, C., Walmsley, C.M., Jewell, P.R., \& Baudry, A. 1988,
A\&A, 199, L5

Jorissen, A., Smith, V.V., \& Lambert, D.L. 1992, A\&A, 261, 164

Kessler M.F., et al.\ 1996, A\&A, 315, L27

Muenter, J.S. 1972, J.\ Chem.\ Phys., 56, 5409

Neufeld, D.A., \& Green, S. 1994, ApJ, 432, 158

Nolt, I.G., et al.\ 1987, J.\ Molec.\ Spectrosc., 125, 274

Ohishi, M. 1997, in IAU Symposium 178, {\it Molecules in Astrophysics: 
Probes and Processes}, ed. E.F. van Dishoeck (Dordrecht: Kluwer), p.\ 61

Rybicki, G.B., \& Hummer, D.G. 1991, A \&A, 245, 171

Schilke, P., Phillips, T.G., \& Wang, N. 1995, ApJ, 441, 334

Stevens, P.S., Brune, W.H., \& Anderson, J.G. 1989, 
  J.\ Phys.\ Chem., 93, 4068

Trams, N.R., Clegg, P.E. \& Swinyard B.M. 1996, {\it Addendum to the LWS
Observers Manual}, Version 1.0, SAI/96-166/Dc (5 August 1996) 
(http://isowww.estec.esa.nl/manuals/LWS\_add/lws\_add2.html)

Toth, R.A. 1991, J.\ Opt.\ Soc.\ Am., 8, 2236

van Dishoeck, E.F., \& Helmich, F.P. 1996, A\&A, 315, L177

Wurzberg, E., \& Houston, P.L. 1980, J.\ Chem.\ Phys., 72, 4811

Ziurys, L.M., Apponi, A.J., \& Phillips, T.G. 1994, ApJ, 433, 729

Zmuidzinas, J., Blake, G.A., Carlstrom, J., Keene, J.,
  \& Miller, D. 1995a, ApJ, 447, L125

Zmuidzinas, J., Blake, G.A., Carlstrom, J., Keene, J., 
  Miller, D., Schilke, P., and Ugras, N.G. 1995b, 
  in {\it Proceedings of the Airborne Astronomy Symposium 
  on the Galactic Ecosystem: From Gas to Stars to Dust}, 
  ed.\ M.R.\ Haas, J.A.\ Davidson \& E.F.\ Erickson
  (San Francisco: ASP), p.\ 33}

\vfill\eject
\centerline{\bf Figure Captions}
\vskip 0.1 true in 
{\hoffset -20pt
\parindent -20pt
Fig. 1 -- ISO LWS Fabry-Perot spectrum of the HF $J=2-1$ line
toward Sgr B2.   The ISO beam, of approximate diameter
$70^{\prime\prime}$ FWHM, was centered midway between
Sgr B2(M) and Sgr B2(N), at coordinates $\rm \alpha=17h\,47m\,20.0s$,
$\delta=-28^\circ \,22^\prime\,41.3 ^{\prime\prime}$ (J2000). 
The flux has been normalized relative to the 
continuum flux of $4.8 \times 10^4$~Jy.  The LSR velocity plotted
on the horizontal axis applies to the  HF $J=2-1$ line at
121.6973 $\mu$m (rest wavelength).  
The solid line shows the fit obtained from the fitting procedure
described in the text.
The location
of the $\rm H_2^{16}O$ $4_{32}-4_{23}$ line in that fit is
shown for the case $v_{\rm LSR} = \rm 67\, km \, s^{-1}$. 

Fig. 2 -- Fractional level populations for rotational states
of HF, as a function of radius, obtained using 
the excitation model described in the text.}

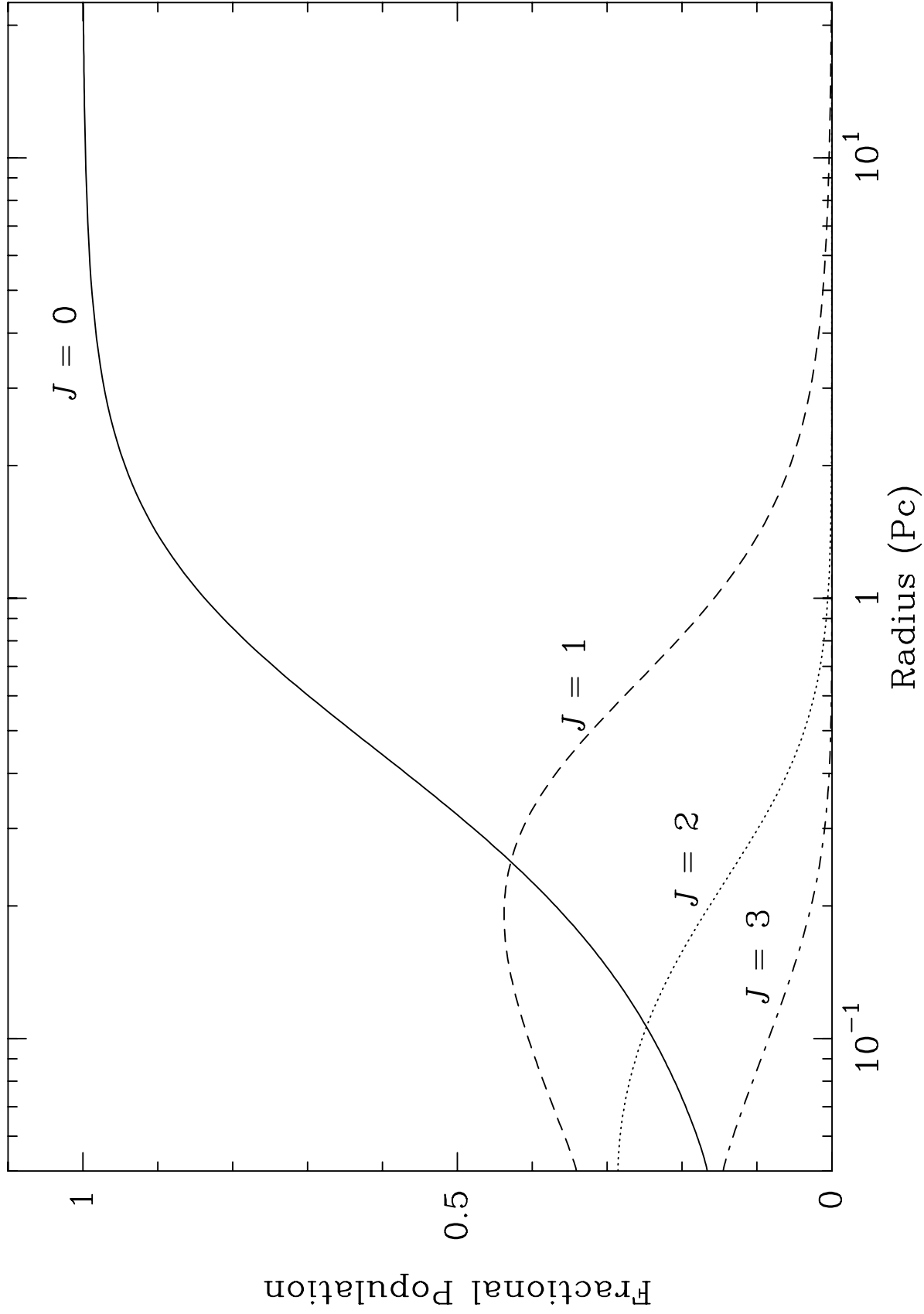
\end{document}